\documentclass[11pt,a4paper]{article}
\newcommand{\preprintassetpath}{}
\IfFileExists{arxiv/source/preprint.sty}{%
  \renewcommand{\preprintassetpath}{arxiv/source/}%
}{\IfFileExists{source/preprint.sty}{%
  \renewcommand{\preprintassetpath}{source/}%
}{}}
\makeatletter
\def\input@path{{arxiv/source/}{source/}}
\makeatother
\usepackage{preprint}
\graphicspath{{arxiv/source/}{source/}}
\definecolor{score0}{HTML}{FAFAFB}
\definecolor{score1}{HTML}{EEEFF3}
\definecolor{score2}{HTML}{E1E3EA}
\definecolor{score3}{HTML}{D2D5E0}
\definecolor{score4}{HTML}{C2C5D4}
\definecolor{score5}{HTML}{B0B2C7}
\definecolor{score6}{HTML}{9E9FBA}
\definecolor{score7}{HTML}{8B8CAD}
\newcommand{\scorecell}[2]{\cellcolor{score#1}\textcolor{black}{#2}}
\title{\bfseries Encoded but Exposed: Language Embeddings can Reveal More Than They Should in Recommender Systems}
\author{Linda Beková \qquad Rodrigo Alves\\[0.5em]
  \small Czech Technical University in Prague\\
  \small Prague, Czech Republic\\[0.25em]
  \small \texttt{bekovlin@fit.cvut.cz} \quad \texttt{rodrigo.alves@fit.cvut.cz}}
\date{}
\hypersetup{pdftitle={Encoded but Exposed: Language Embeddings can Reveal More Than They Should in Recommender Systems},pdfauthor={Linda Beková and Rodrigo Alves}}

\begin{document}
\maketitle
\begin{center}
\begin{minipage}{0.95\textwidth}
\small\centering\itshape
Accepted for presentation at the 3rd Workshop on Evaluating and Applying Recommender Systems with Large Language Models (EARL 2026), co-located with the 35th ACM International Conference on Information and Knowledge Management (CIKM 2026), November 8, 2026, Rome, Italy.
\end{minipage}
\end{center}

\begin{abstract}
Language embeddings are increasingly used in recommenders to represent item content, but their privacy risks remain poorly understood. We study whether item embeddings can reveal item identities when an adversary has access to auxiliary item text from external sources. We show that a simple cosine-similarity attack can recover item identities with high mean reciprocal rank across language models and candidate-set settings on two real-world datasets. Leakage is strongest when the adversary's partial information overlaps with the victim representation, but it also persists through semantic alignment between independently sourced descriptions. We further show that item leakage can propagate to user re-identification from sparse ratings. Finally, we evaluate obfuscated descriptions as a mitigation and find a clear leakage--utility tradeoff.
\end{abstract}

\section{Introduction}

\paragraph{Motivation} Recommender Systems (RSs) rely on effective representations of users and items to capture preferences~\cite{jeunen2022embarrassingly}, intent~\cite{li2021intention}, and contextual signals~\cite{mateos2024systematic,alves2021burst}. Recent advances in language models have made it possible to derive richer representations from natural language data, such as item descriptions, reviews, and queries~\cite{zhao2024recommender}. Consequently, natural language embeddings are increasingly used in both research~\cite{le2025enhancing,vats2025exploring} and industrial settings~\cite{kasalicky2025future,ni2026treebridge,cui2022m6} to support more accurate and context-aware recommendations.

Besides improving downstream recommendation tasks, language embeddings are \emph{often perceived} as privacy-preserving because they are high-dimensional and not directly interpretable by humans. For instance, embedding representations of users and items are frequently shared with third-party service providers~\cite{gao2021cross,hu2020privnet,wu2025aegis}. However, this assumption can be particularly fragile in recommendation system scenarios~\cite{calandrino2011you,chang2010towards}. Although a potential adversary may not have access to the original corpus used to generate the embeddings, domain knowledge and the abundance of publicly available item data~\cite{harper2015movielens,hou2024bridging,wu2020mind}  can make attacks more feasible than in settings where the underlying information is entirely internal or difficult to obtain. Therefore, even if an attacker does not have access to the exact movie plots used by a streaming provider to generate such embeddings, publicly available movie datasets~\cite{oghina2012predicting,sayifullah2026exploratory} can be used to construct a similar corpus, enabling effective inference attacks. Our work focuses on studying the robustness of item embeddings generated by different language models in recommender settings by examining whether it is possible to identify which items correspond to a given set of embeddings.

\paragraph{Related Literature}

Song and Raghunathan~\cite{song2020information} provide a \textbf{taxonomy of privacy attacks against embeddings}, showing that the typical goals are to infer private attributes from embeddings (e.g., the title or genre of a movie)~\citep{li2023sentence,huang2024transferable,song2019overlearning}, determine whether specific data or contexts were used during training~\citep{mahloujifar2021membership,mattern2023membership}, or reconstruct the original input text from its representation~\citep{huang2024transferable,morris2023text}. Our work is more closely related to the first category. In this setting, the adversary is assumed to have access to the embedding function and to an auxiliary labeled dataset that contains labels of a discrete sensitive attribute. The attacker then learns a classifier over the embedding space in order to infer sensitive attributes directly from the embedding representations.

\textbf{Attribute inference} from embedding-based models has been studied as a privacy risk across domains. Prior work shows that text embeddings can encode sensitive attributes even when the original text is unavailable~\cite{zhu2024understanding}, and similar leakage has been observed in computer vision embeddings designed for utility tasks~\cite{abedi2024attribute}. More broadly,~\cite{song2019overlearning} show that models can encode information beyond the training objective, a phenomenon they call \emph{overlearning}. In RSs, this risk has mainly been studied for user representations: transferred~\cite{hu2020privnet} or graph-based~\cite{zhang2021graph} recommender embeddings can leak private user attributes, motivating recent attribute-unlearning and in-training defenses~\cite{wu2025aegis,feng2025raid,li2026fedau2}. In contrast, our work shifts the focus to item-level privacy, treating item identity as a discrete sensitive attribute inferred from item embeddings.

Another line of work in RS privacy and security studies whether \textbf{users or user-item interactions can be de-anonymized} from recommendation data. Early work showed that sparse rating patterns are often distinctive enough to re-identify users in anonymized datasets~\cite{narayanan2008robust}. Subsequent studies demonstrated that even without explicit demographics, rating behavior can reveal sensitive user attributes: for example,~\cite{weinsberg2012blurme} infers user gender from ratings using only a small set of users with disclosed demographic labels. More recent work extends this perspective to membership inference, asking whether a user or a specific user--item interaction was part of a recommender's training data~\cite{zhang2021membership,zhong2024interaction}, and to cross-domain record linkage, where rating patterns are used to link user identities across datasets from different domains~\cite{torres2024anonymizing}. Our work connects to this direction by showing that item-level identity leakage can propagate to user-level privacy risks.

\paragraph{Contributions.} Our contributions are as follows: we (i) formulate item-identity inference from exposed language embeddings in RSs; (ii) show substantial leakage across datasets, candidate settings, and models; (iii) study obfuscation and its leakage--utility tradeoff; and (iv) show that item leakage can propagate to user re-identification.

\section{Method}

\paragraph{Problem Definition}
We consider a recommender in which users interact with items, such as movies, books, or products. Let $\mathcal{U}$ denote the set of users and $\mathcal{I}$ denote the set of items. Each item $i \in \mathcal{I}$ is associated with textual information $x_i$ (e.g., movie description) and a discrete sensitive attribute $y_i$ (resp., movie title). We denote the underlying victim's corpus by $\mathcal{D}=\{(x_i,y_i): i \in \mathcal{I}\}$, which is not directly observed by the adversary. Instead, the recommendation provider may expose the semantic information only through embeddings $z_i = \phi(x_i) \in \mathbb{R}^d$, where $\phi$ is an embedding function. In our setting, we assume that $\phi$ is a pretrained language model.

Aligned with prior work~\cite{song2020information}, we assume that the adversary has access to $\phi$ and to an auxiliary labeled dataset $\widebar{\mathcal{D}}=\{(\widebar{x}_j,\widebar{y}_j)\}_{j=1}^{n}$. This is particularly realistic in RSs settings, where the item domain is often known and large amounts of publicly available item information can be used to construct $\widebar{\mathcal{D}}$. The adversary embeds auxiliary data using $\phi$ and builds a classifier over the embedding space to infer the sensitive attribute from exposed embeddings. More formally, let $f_{\widebar{\mathcal{D}}}$ be the adversary classifier built using $\widebar{\mathcal{D}}$, such that
\begin{align}
    \hat{y}_i = f_{\widebar{\mathcal{D}}}(\phi(x_i)),
\end{align}
where $\hat{y}_i$ denotes the inferred item identity associated with the embedding of item $i$. We aim to quantify identity leakage as the adversary's ability to recover $y_i$ from the exposed embedding $\phi(x_i)$, considering embeddings produced by different language models.

\paragraph{Adversary Scenarios}
In many RS domains, item text data is available outside the victim platform through, for instance, public catalogues, aggregators, or even AI-generated descriptions. For instance, movies can often be linked to resources such as IMDb or TMDb, which openly provide titles, plots, descriptions, and more. Such sources may allow an adversary to build an auxiliary catalogue~$\widebar{\mathcal{I}}$ without access to the victim's internal corpus. We distinguish two cases according to the overlap with the victim item set $\mathcal{I}$:

\begin{itemize}
    \item \textbf{Closed candidate set:} $\mathcal{I} = \widebar{\mathcal{I}}$; the adversary matches exposed embeddings to known item identities.
    \item \textbf{Open candidate set:} $\mathcal{I} \subsetneq \widebar{\mathcal{I}}$; the adversary retrieves victim items from a larger catalogue with distractors.
\end{itemize}
The open setting is strictly more challenging, as it adds candidate selection to the matching problem. Both cases are realistic: in some domains the relevant catalogue is nearly complete, while in others the attacker may only have access to a broader external catalogue whose overlap with the victim platform is unknown.

\paragraph{Item Identification Pipeline}

Given access to the auxiliary dataset $\widebar{\mathcal{D}}$,  the adversary can construct embeddings that are directly comparable to the leaked victim embeddings. This allows the item-identification problem to be formulated as a candidate-ranking problem in the embedding space. In particular, for each candidate item $(\widebar{x}_j,\widebar{y}_j) \in \widebar{\mathcal{D}}$, we assume an adversary computes
\begin{align}
\widebar{z}_j = \phi(\widebar{x}_j),
\end{align}
using the same model $\phi$ that generates the victim's embeddings~$z_i$. The adversary then defines a scoring function $g: \mathbb{R}^d \times \mathbb{R}^d \rightarrow \mathbb{R},$ which assigns a compatibility score between a leaked victim embedding and a candidate embedding. This function may be a fixed similarity measure, such as cosine similarity, or a learned model trained on the auxiliary dataset. For each victim embedding $z_i$, the adversary computes
\begin{align}
\pi_i(j)=g(z_i,\widebar{z}_j),
\qquad j \in \widebar{\mathcal{I}},
\end{align}
which induces a ranking over the candidate catalogue. Candidates with higher scores are considered more likely to correspond to the leaked embedding $z_i$. The resulting ranking can then be used to recover the victim item identity or to narrow the search space to a small set of plausible candidates. In this view, the scoring function $g$ provides a ranking-based formulation of $f_{\overline{\mathcal{D}}}$. Specifically, $f_{\overline{\mathcal{D}}}$ returns the highest-scoring candidate identity,

\begin{align}
f_{\overline{\mathcal{D}}}(z_i)
=
\overline{y}_{j^\ast},
\qquad
j^\ast = \arg\max_{j \in \overline{\mathcal{I}}} g(z_i,\overline{z}_j).
\end{align}

\paragraph{\textbf{Remark on the Known Embedding Function.}}
We assume that the adversary knows and can use the embedding function $\phi$, a common assumption in embedding privacy attacks~\cite{song2020information}. This allows us to isolate item-identity leakage given a shared embedding space. We also conducted preliminary autoencoder experiments to identify an unknown encoder: for each candidate model with the same output dimensionality as the leaked embeddings, we trained an autoencoder on embeddings of auxiliary texts and compared its reconstruction error on the victim embeddings. Across the evaluated movie and book settings, the autoencoder trained on embeddings from the victim's encoder consistently achieved the lowest reconstruction error. These results suggest that the encoder can be identified within the tested candidate set; they do not establish identification of arbitrary models or transformed embeddings.

\section{Experiments}

\paragraph{Experimental Setup and Reproducibility Details}
We instantiate the embedding function $\phi$ using six sentence-transformer models that are popular on Hugging Face and commonly used as off-the-shelf text encoders in the RS community:

\begin{itemize}
    \item \textbf{MiniLM-L6}: \texttt{all-MiniLM-L6-v2};
    \item \textbf{MiniLM-L3}: \texttt{paraphrase-MiniLM-L3-v2};
    \item \textbf{mMiniLM-L12}: \texttt{paraphrase-multilingual-MiniLM-L12-v2};
    \item \textbf{DistilRoBERTa}: \texttt{all-distilroberta-v1};
    \item \textbf{MPNet-QA}: \texttt{multi-qa-mpnet-base-dot-v1};
    \item \textbf{ALBERT-small}: \texttt{paraphrase-albert-small-v2};
\end{itemize}
Thus, for the first three models, $z_i \in \mathbb{R}^{384}$; for the last three, $z_i \in \mathbb{R}^{768}$. We use two established benchmark datasets from different domains (movies and books, respectively): MovieLens~32M~\cite{harper2015movielens} and GoodReads~\cite{wan2018item}. For each dataset, we retain the top $10$K most popular items. For MovieLens, we use the victim corpus $x_i$ collected in~\cite{kamath_movielens32m_sids}, while, for GoodReads, we use the metadata provided by the original dataset. We consider two victim-corpus  settings: \textbf{($\mathbf{D}$)} item descriptions, corresponding to movie plots or book synopses, and \textbf{($\mathbf{T+D}$)} titles combined with descriptions. For the adversarial corpus $\bar{x}_i$, we consider two settings: \textbf{($\widebar{\mathbf{D}}$)} descriptions generated using \texttt{gpt-oss-20b}, and \textbf{($\widebar{\mathbf{T}}$)} titles only. Importantly, note that $D$ and $\widebar{D}$ come from different sources, so description-based identification reflects semantic similarity rather than text reuse.

To emulate the \emph{closed candidate set}, we vary the number of items of the victim dataset, considering $|{\mathcal{I}}| = |\bar{\mathcal{I}}| \in \{1000,3000,5000\}$. In contrast, for the \emph{open candidate set}, we fix $|\bar{\mathcal{I}}| = 1000$ and vary $|\mathcal{I}| \in \{2000,3000,4000\}$ to capture different proportions between the leaked list and the full item set. For each combination of victim corpus, adversarial corpus, $|\mathcal{I}|$, and $|\widebar{\mathcal{I}}|$, we sample from the original dataset items uniformly at random according to the corresponding candidate-set setting.
We instantiate the scoring function as cosine similarity, $g(z_i,\widebar{z}_j)=z_i^\top \widebar{z}_j / (\|z_i\|\,\|\widebar{z}_j\|)$. This non-parametric choice allows us to measure identity leakage directly from the embedding space, without adding capacity through a learned attacker.
As evaluation metric, we report mean reciprocal rank (MRR), a widely used retrieval measure defined as the average inverse rank of the correct item: higher values indicate better ranking, with $1$ corresponding to rank $1$ and $0.5$ roughly to rank $2$ on average.

\begin{table}[t]
\caption{Item-identification MRR across datasets, candidate settings, embedding models, victim corpora, and adversarial corpora.}
\centering
\setlength{\tabcolsep}{3pt}
\renewcommand{\arraystretch}{1.15}
\setlength{\aboverulesep}{0.3ex}
\setlength{\belowrulesep}{0.3ex}

\resizebox{\textwidth}{!}{
\begin{tabular}{c c c c
ccc !{\color{white}\vrule width 2pt}
ccc !{\color{white}\vrule width 2pt}
ccc !{\color{white}\vrule width 2pt}
ccc !{\color{white}\vrule width 2pt}
ccc !{\color{white}\vrule width 2pt}
ccc}
\toprule

& & & & \multicolumn{3}{c}{\textbf{MiniLM-L6}}
& \multicolumn{3}{c}{\textbf{MiniLM-L3}}
& \multicolumn{3}{c}{\textbf{mMiniLM-L12}}
& \multicolumn{3}{c}{\textbf{DistilRoBERTa}}
& \multicolumn{3}{c}{\textbf{MPNet-QA}}
& \multicolumn{3}{c}{\textbf{ALBERT-small}} \\
\cmidrule(lr){5-7}
\cmidrule(lr){8-10}
\cmidrule(lr){11-13}
\cmidrule(lr){14-16}
\cmidrule(lr){17-19}
\cmidrule(lr){20-22}

& & & \textbf{A/V}
& $T+D$ & $D$ & $\tilde{D}$
& $T+D$ & $D$ & $\tilde{D}$
& $T+D$ & $D$ & $\tilde{D}$
& $T+D$ & $D$ & $\tilde{D}$
& $T+D$ & $D$ & $\tilde{D}$
& $T+D$ & $D$ & $\tilde{D}$\\
\midrule

\multirow{12}{*}[-1.5ex]{\rotatebox{90}{{MovieLens}}}
& \multirow{6}{*}[-0.8ex]{\rotatebox{90}{Closed Set}}
& \multirow{2}{*}{1000}
& $\widebar{T}$
& \scorecell{7}{1.00} & \scorecell{7}{0.91} & \scorecell{0}{0.15} & \scorecell{7}{0.97} & \scorecell{6}{0.78} & \scorecell{0}{0.12} & \scorecell{7}{0.95} & \scorecell{6}{0.79} & \scorecell{0}{0.14} & \scorecell{7}{1.00} & \scorecell{7}{0.92} & \scorecell{0}{0.17} & \scorecell{7}{0.98} & \scorecell{7}{0.89} & \scorecell{0}{0.12} & \scorecell{7}{0.96} & \scorecell{6}{0.78} & \scorecell{0}{0.12}\\
& & & $\widebar{D}$
& \scorecell{6}{0.80} & \scorecell{6}{0.79} & \scorecell{3}{0.46} & \scorecell{6}{0.77} & \scorecell{5}{0.73} & \scorecell{3}{0.51} & \scorecell{5}{0.75} & \scorecell{5}{0.72} & \scorecell{3}{0.51} & \scorecell{6}{0.80} & \scorecell{6}{0.78} & \scorecell{3}{0.45} & \scorecell{6}{0.81} & \scorecell{6}{0.82} & \scorecell{3}{0.43} & \scorecell{5}{0.73} & \scorecell{5}{0.70} & \scorecell{3}{0.47}\\
\cmidrule(lr){3-22}

& & \multirow{2}{*}{3000}
& $\widebar{T}$
& \scorecell{7}{0.99} & \scorecell{6}{0.88} & \scorecell{0}{0.10} & \scorecell{7}{0.95} & \scorecell{5}{0.72} & \scorecell{0}{0.08} & \scorecell{7}{0.92} & \scorecell{5}{0.74} & \scorecell{0}{0.09} & \scorecell{7}{0.99} & \scorecell{7}{0.89} & \scorecell{0}{0.11} & \scorecell{7}{0.96} & \scorecell{6}{0.85} & \scorecell{0}{0.08} & \scorecell{7}{0.95} & \scorecell{5}{0.73} & \scorecell{0}{0.08} \\
& & & $\widebar{D}$
& \scorecell{5}{0.75} & \scorecell{5}{0.74} & \scorecell{2}{0.37} & \scorecell{5}{0.71} & \scorecell{5}{0.67} & \scorecell{2}{0.41} & \scorecell{5}{0.69} & \scorecell{5}{0.66} & \scorecell{2}{0.41} & \scorecell{5}{0.75} & \scorecell{5}{0.73} & \scorecell{2}{0.36} & \scorecell{5}{0.76} & \scorecell{5}{0.76} & \scorecell{2}{0.35} & \scorecell{5}{0.68} & \scorecell{4}{0.64} & \scorecell{2}{0.37}  \\
\cmidrule(lr){3-22}

& & \multirow{2}{*}{5000}
& $\widebar{T}$
& \scorecell{7}{0.99} & \scorecell{6}{0.86} & \scorecell{0}{0.08} & \scorecell{7}{0.94} & \scorecell{5}{0.69} & \scorecell{0}{0.06} & \scorecell{7}{0.91} & \scorecell{5}{0.71} & \scorecell{0}{0.08} & \scorecell{7}{0.99} & \scorecell{6}{0.88} & \scorecell{0}{0.09} & \scorecell{7}{0.95} & \scorecell{6}{0.83} & \scorecell{0}{0.06} & \scorecell{7}{0.93} & \scorecell{5}{0.71} & \scorecell{0}{0.06}\\
& & & $\widebar{D}$
& \scorecell{5}{0.73} & \scorecell{5}{0.71} & \scorecell{2}{0.33} & \scorecell{5}{0.68} & \scorecell{4}{0.64} & \scorecell{2}{0.37} & \scorecell{5}{0.66} & \scorecell{4}{0.63} & \scorecell{2}{0.37} & \scorecell{5}{0.73} & \scorecell{5}{0.70} & \scorecell{2}{0.32} & \scorecell{5}{0.73} & \scorecell{5}{0.73} & \scorecell{2}{0.32} & \scorecell{5}{0.65} & \scorecell{4}{0.61} & \scorecell{2}{0.33} \\
\cmidrule(lr){2-22}

& \multirow{6}{*}[-0.8ex]{\rotatebox{90}{Open Set}}
& \multirow{2}{*}{2000}
& $\widebar{T}$
& \scorecell{7}{0.99} & \scorecell{7}{0.89} & \scorecell{0}{0.12} & \scorecell{7}{0.96} & \scorecell{5}{0.74} & \scorecell{0}{0.09} & \scorecell{7}{0.93} & \scorecell{5}{0.76} & \scorecell{0}{0.11} & \scorecell{7}{1.00} & \scorecell{7}{0.90} & \scorecell{0}{0.13} & \scorecell{7}{0.97} & \scorecell{6}{0.87} & \scorecell{0}{0.09} & \scorecell{7}{0.95} & \scorecell{5}{0.75} & \scorecell{0}{0.09} \\
& & & $\widebar{D}$
& \scorecell{6}{0.77} & \scorecell{5}{0.76} & \scorecell{2}{0.41} & \scorecell{5}{0.73} & \scorecell{5}{0.69} & \scorecell{3}{0.45} & \scorecell{5}{0.71} & \scorecell{5}{0.68} & \scorecell{3}{0.44} & \scorecell{6}{0.77} & \scorecell{5}{0.75} & \scorecell{2}{0.39} & \scorecell{6}{0.77} & \scorecell{6}{0.79} & \scorecell{2}{0.38} & \scorecell{5}{0.70} & \scorecell{5}{0.66} & \scorecell{2}{0.40} \\
\cmidrule(lr){3-22}

& & \multirow{2}{*}{3000}
& $\widebar{T}$
& \scorecell{7}{0.99} & \scorecell{6}{0.88} & \scorecell{0}{0.10} & \scorecell{7}{0.95} & \scorecell{5}{0.72} & \scorecell{0}{0.08} & \scorecell{7}{0.92} & \scorecell{5}{0.74} & \scorecell{0}{0.10} & \scorecell{7}{1.00} & \scorecell{7}{0.89} & \scorecell{0}{0.11} & \scorecell{7}{0.96} & \scorecell{6}{0.85} & \scorecell{0}{0.08} & \scorecell{7}{0.95} & \scorecell{5}{0.73} & \scorecell{0}{0.08} \\
& & & $\widebar{D}$
& \scorecell{5}{0.75} & \scorecell{5}{0.74} & \scorecell{2}{0.37} & \scorecell{5}{0.71} & \scorecell{5}{0.67} & \scorecell{2}{0.41} & \scorecell{5}{0.69} & \scorecell{5}{0.66} & \scorecell{2}{0.41} & \scorecell{5}{0.75} & \scorecell{5}{0.73} & \scorecell{2}{0.36} & \scorecell{5}{0.76} & \scorecell{6}{0.76} & \scorecell{2}{0.35} & \scorecell{5}{0.68} & \scorecell{4}{0.64} & \scorecell{2}{0.37} \\
\cmidrule(lr){3-22}

& & \multirow{2}{*}{4000}
& $\widebar{T}$
& \scorecell{7}{0.99} & \scorecell{6}{0.87} & \scorecell{0}{0.09} & \scorecell{7}{0.95} & \scorecell{5}{0.71} & \scorecell{0}{0.07} & \scorecell{7}{0.91} & \scorecell{5}{0.72} & \scorecell{0}{0.08} & \scorecell{7}{0.99} & \scorecell{7}{0.88} & \scorecell{0}{0.10} & \scorecell{7}{0.96} & \scorecell{6}{0.84} & \scorecell{0}{0.07} & \scorecell{7}{0.94} & \scorecell{5}{0.72} & \scorecell{0}{0.07} \\
& & & $\widebar{D}$
& \scorecell{5}{0.74} & \scorecell{5}{0.72} & \scorecell{2}{0.35} & \scorecell{5}{0.69} & \scorecell{5}{0.65} & \scorecell{2}{0.39} & \scorecell{5}{0.67} & \scorecell{4}{0.64} & \scorecell{2}{0.39} & \scorecell{5}{0.74} & \scorecell{5}{0.71} & \scorecell{2}{0.34} & \scorecell{5}{0.74} & \scorecell{5}{0.75} & \scorecell{2}{0.33} & \scorecell{5}{0.67} & \scorecell{4}{0.62} & \scorecell{2}{0.35} \\

\midrule

\multirow{12}{*}[-1.5ex]{\rotatebox{90}{GoodReads}}
& \multirow{6}{*}[-0.8ex]{\rotatebox{90}{Closed Set}}
& \multirow{2}{*}{1000}
& $\widebar{T}$
& \scorecell{7}{0.95} & \scorecell{5}{0.68} & \scorecell{2}{0.37} & \scorecell{6}{0.87} & \scorecell{4}{0.54} & \scorecell{1}{0.28} & \scorecell{5}{0.73} & \scorecell{3}{0.45} & \scorecell{2}{0.34} & \scorecell{7}{0.95} & \scorecell{5}{0.69} & \scorecell{2}{0.41} & \scorecell{7}{0.91} & \scorecell{5}{0.66} & \scorecell{2}{0.34} & \scorecell{7}{0.90} & \scorecell{4}{0.59} & \scorecell{2}{0.31} \\
& & & $\widebar{D}$
& \scorecell{7}{0.92} & \scorecell{5}{0.71} & \scorecell{3}{0.48} & \scorecell{6}{0.77} & \scorecell{4}{0.55} & \scorecell{3}{0.44} & \scorecell{4}{0.60} & \scorecell{3}{0.44} & \scorecell{3}{0.46} & \scorecell{7}{0.93} & \scorecell{5}{0.71} & \scorecell{3}{0.51} & \scorecell{6}{0.87} & \scorecell{5}{0.67} & \scorecell{3}{0.44} & \scorecell{6}{0.84} & \scorecell{4}{0.61} & \scorecell{3}{0.43} \\
\cmidrule(lr){3-22}

& & \multirow{2}{*}{3000}
& $\widebar{T}$
& \scorecell{7}{0.92} & \scorecell{4}{0.61} & \scorecell{2}{0.31} & \scorecell{6}{0.83} & \scorecell{3}{0.47} & \scorecell{1}{0.21} & \scorecell{5}{0.67} & \scorecell{2}{0.38} & \scorecell{1}{0.27} & \scorecell{7}{0.92} & \scorecell{4}{0.63} & \scorecell{2}{0.34} & \scorecell{6}{0.87} & \scorecell{4}{0.60} & \scorecell{1}{0.28} & \scorecell{6}{0.87} & \scorecell{3}{0.52} & \scorecell{1}{0.25} \\
& & & $\widebar{D}$
& \scorecell{7}{0.88} & \scorecell{4}{0.63} & \scorecell{2}{0.40} & \scorecell{5}{0.71} & \scorecell{3}{0.47} & \scorecell{2}{0.35} & \scorecell{3}{0.52} & \scorecell{2}{0.36} & \scorecell{2}{0.38} & \scorecell{7}{0.89} & \scorecell{4}{0.64} & \scorecell{3}{0.43} & \scorecell{6}{0.82} & \scorecell{4}{0.60} & \scorecell{2}{0.36} & \scorecell{6}{0.79} & \scorecell{4}{0.53} & \scorecell{2}{0.35} \\
\cmidrule(lr){3-22}

& & \multirow{2}{*}{5000}
& $\widebar{T}$
& \scorecell{7}{0.90} & \scorecell{4}{0.58} & \scorecell{1}{0.28} & \scorecell{6}{0.81} & \scorecell{3}{0.44} & \scorecell{1}{0.19} & \scorecell{4}{0.64} & \scorecell{2}{0.34} & \scorecell{1}{0.25} & \scorecell{7}{0.90} & \scorecell{4}{0.59} & \scorecell{2}{0.31} & \scorecell{6}{0.85} & \scorecell{4}{0.56} & \scorecell{1}{0.25} & \scorecell{6}{0.86} & \scorecell{3}{0.49} & \scorecell{1}{0.23} \\
& & & $\widebar{D}$
& \scorecell{6}{0.86} & \scorecell{4}{0.60} & \scorecell{2}{0.36} & \scorecell{5}{0.67} & \scorecell{3}{0.43} & \scorecell{2}{0.32} & \scorecell{3}{0.49} & \scorecell{2}{0.33} & \scorecell{2}{0.34} & \scorecell{6}{0.86} & \scorecell{4}{0.61} & \scorecell{2}{0.39} & \scorecell{6}{0.79} & \scorecell{4}{0.56} & \scorecell{2}{0.33} & \scorecell{5}{0.76} & \scorecell{3}{0.50} & \scorecell{2}{0.32} \\
\cmidrule(lr){2-22}

& \multirow{6}{*}[-0.8ex]{\rotatebox{90}{Open Set}}
& \multirow{2}{*}{2000}
& $\widebar{T}$
& \scorecell{7}{0.93} & \scorecell{4}{0.64} & \scorecell{2}{0.33} & \scorecell{6}{0.84} & \scorecell{3}{0.50} & \scorecell{1}{0.24} & \scorecell{5}{0.69} & \scorecell{2}{0.40} & \scorecell{2}{0.30} & \scorecell{7}{0.93} & \scorecell{5}{0.65} & \scorecell{2}{0.36} & \scorecell{7}{0.88} & \scorecell{4}{0.62} & \scorecell{2}{0.30} & \scorecell{7}{0.89} & \scorecell{4}{0.55} & \scorecell{1}{0.27} \\
& & & $\widebar{D}$
& \scorecell{7}{0.90} & \scorecell{5}{0.66} & \scorecell{3}{0.43} & \scorecell{5}{0.73} & \scorecell{3}{0.50} & \scorecell{2}{0.38} & \scorecell{4}{0.55} & \scorecell{2}{0.39} & \scorecell{2}{0.41} & \scorecell{7}{0.90} & \scorecell{5}{0.67} & \scorecell{3}{0.46} & \scorecell{6}{0.84} & \scorecell{4}{0.62} & \scorecell{2}{0.39} & \scorecell{6}{0.81} & \scorecell{4}{0.56} & \scorecell{2}{0.38} \\
\cmidrule(lr){3-22}

& & \multirow{2}{*}{3000}
& $\widebar{T}$
& \scorecell{7}{0.92} & \scorecell{4}{0.62} & \scorecell{2}{0.31} & \scorecell{6}{0.83} & \scorecell{3}{0.47} & \scorecell{1}{0.21} & \scorecell{5}{0.67} & \scorecell{2}{0.38} & \scorecell{1}{0.27} & \scorecell{7}{0.92} & \scorecell{4}{0.62} & \scorecell{2}{0.34} & \scorecell{6}{0.87} & \scorecell{4}{0.60} & \scorecell{1}{0.28} & \scorecell{6}{0.87} & \scorecell{3}{0.52} & \scorecell{1}{0.25} \\
& & & $\widebar{D}$
& \scorecell{7}{0.88} & \scorecell{4}{0.63} & \scorecell{2}{0.40} & \scorecell{5}{0.71} & \scorecell{3}{0.47} & \scorecell{2}{0.35} & \scorecell{3}{0.53} & \scorecell{2}{0.36} & \scorecell{2}{0.38} & \scorecell{7}{0.89} & \scorecell{4}{0.64} & \scorecell{3}{0.42} & \scorecell{6}{0.82} & \scorecell{4}{0.60} & \scorecell{2}{0.36} & \scorecell{6}{0.79} & \scorecell{4}{0.54} & \scorecell{2}{0.35} \\
\cmidrule(lr){3-22}

& & \multirow{2}{*}{4000}
& $\widebar{T}$
& \scorecell{7}{0.91} & \scorecell{4}{0.60} & \scorecell{1}{0.29} & \scorecell{6}{0.82} & \scorecell{3}{0.45} & \scorecell{1}{0.20} & \scorecell{5}{0.66} & \scorecell{2}{0.36} & \scorecell{1}{0.26} & \scorecell{7}{0.91} & \scorecell{4}{0.61} & \scorecell{2}{0.32} & \scorecell{6}{0.86} & \scorecell{4}{0.58} & \scorecell{1}{0.27} & \scorecell{6}{0.87} & \scorecell{3}{0.50} & \scorecell{1}{0.24} \\
& & & $\widebar{D}$
& \scorecell{6}{0.87} & \scorecell{4}{0.62} & \scorecell{2}{0.38} & \scorecell{5}{0.69} & \scorecell{3}{0.44} & \scorecell{2}{0.33} & \scorecell{3}{0.51} & \scorecell{2}{0.34} & \scorecell{2}{0.36} & \scorecell{6}{0.87} & \scorecell{4}{0.62} & \scorecell{2}{0.41} & \scorecell{6}{0.80} & \scorecell{4}{0.58} & \scorecell{2}{0.34} & \scorecell{6}{0.77} & \scorecell{3}{0.51} & \scorecell{2}{0.33} \\

\bottomrule
\end{tabular}
}

\label{tab:results_transposed_domain_no_fin}
\end{table}

\paragraph{Results.} Table~\ref{tab:results_transposed_domain_no_fin} shows that item embeddings exposed to the proposed attack can reveal substantial identity information across both datasets, candidate-set settings, and embedding models. Each entry is averaged over $100$ random samples of $\mathcal{I}$ and $\widebar{\mathcal{I}}$. Leakage is strongest when the victim corpus contains both title and description ($T+D$). In this setting, title-only auxiliary information ($\widebar{T}$) often yields high MRR, indicating that \emph{partial textual information available to the adversary} can be sufficient for accurate item identification in the evaluated candidate-set settings. The attack remains effective even when victim embeddings are computed from descriptions only (${D}$). This indicates that item identity can be recovered not only from explicit information overlap, but also from semantic alignment. Moreover, MRR decreases only modestly as the candidate set grows in both the closed-set and open-set settings, showing that leakage is not limited to small catalogues. The same pattern holds across all six embedding models. Thus, increasing dimensionality alone does not remove item-identity information.

The two domains differ in the type of auxiliary information that is most effective. In MovieLens, title-based attacks are particularly strong, even when the victim embedding is built only from descriptions. This suggests strong lexical or semantic overlap between movie titles and plot summaries (e.g., title mentions or named entities). In GoodReads, title-based attacks remain effective when the victim representation includes titles, but description-based auxiliary information is often competitive, especially against description-only victim embeddings. This suggests that book identity relies more on semantic alignment than exact textual cues.

\begin{figure}[t]
    \centering
    \includegraphics[width=\textwidth]{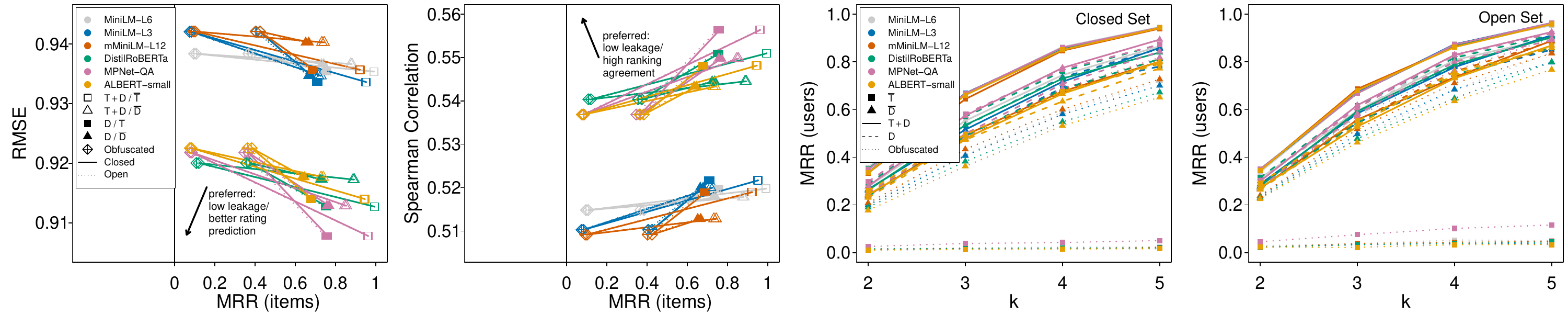}
\caption{Leakage--utility tradeoff and user re-identification results on MovieLens. Left: obfuscation reduces item-identification MRR but degrades rating prediction and ranking agreement; right: sparse ratings enable user re-identification as $k$ increases.}
    \label{fig:obfuscated_users}
\end{figure}

\paragraph{Leakage--Utility Tradeoff.}
As a simple mitigation, the victim could replace the original item text ($T+D$ or $D$) with obfuscated descriptions that preserve general semantics while removing explicit identifiers such as titles, actors, directors, producers, or franchise-specific names. To evaluate this strategy, we construct an obfuscated corpus $\tilde{D}$ for a random sample of $5000$ movies using~\texttt{gpt-oss-20b}. Each description is generated with a prompt that explicitly instructs the model to preserve the general plot semantics while excluding direct identifiers. The results are shown in the $\tilde{D}$ columns of Table~\ref{tab:results_transposed_domain_no_fin}. These results show that obfuscation substantially reduces identity leakage, especially for title-only attacks. In MovieLens, $\widebar{T}$-based MRR drops from very high values under $T+D$ or $D$ to roughly $0.06$--$0.17$ across models and candidate settings, indicating that removing explicit identifiers largely breaks the title-driven matching effect. However, leakage remains non-negligible when the adversary uses auxiliary descriptions ($\widebar{D}$), with MRR typically around $0.32$--$0.51$. Thus, obfuscated plots still preserve semantic information that can support partial item identification. Across candidate settings and embedding models, obfuscation consistently lowers MRR for all six encoders, but no model eliminates leakage.

However, reducing identity leakage may also reduce the semantic expressiveness of the item embeddings. We therefore evaluate, on MovieLens, how obfuscated embeddings affect downstream recommendation quality, a typical task for which such embeddings are designed. To do so, we use an inductive matrix factorization~\cite{chiang2018using,ledent2021orthogonal} model in which the rating matrix is approximated as
\begin{equation}
    R \approx M H,
\end{equation}
where $M$ is a learned low-rank user matrix that captures collaborative-filtering structure, and $H \in \{T, D, \tilde{D}\}$ denotes the item embedding representation. For each combination of candidate setting, embedding model $\phi$, victim corpus $\mathcal{D}$, and adversarial corpus $\widebar{\mathcal{D}}$, we evaluate the resulting representation along two dimensions: privacy leakage, measured by MRR, and downstream recommendation quality. We report RMSE to measure rating-prediction error and Spearman correlation to measure ranking agreement. In Figure~\ref{fig:obfuscated_users}, each point reports the average metric value over the candidate-set sizes $|\mathcal{I}|$ and $|\widebar{\mathcal{I}}|$, while the trajectories show the effect of replacing the original victim representation with the obfuscated corpus $\tilde{D}$ across the remaining experimental parameters. Across parameter combinations, obfuscation consistently reduces MRR, making item identities harder to recover. However, this privacy gain comes with a clear utility cost: $\tilde{D}$ increases RMSE and reduces Spearman correlation, indicating worse rating prediction and weaker preservation of recommendation ranking quality.

\paragraph{User Re-identification.}
Prior work has shown that sparse interaction data can be sufficient to re-identify users under different contexts~\cite{narayanan2008robust,weinsberg2012blurme,zhong2024interaction}. A common re-identification scenario assumes that the adversary observes a few interactions from a target user and attempts to locate the corresponding user in the rating matrix, thereby recovering their full interaction history. We adapt this setting to our item-identification scenario. Consider a target user $u^\star$ for whom the adversary observes a small set of $k$ rated items, $A_{u^\star}=\{(j,r_{u^\star,j})\}_{j=1}^{k}$, where $j$ is an observed item and $r_{u^\star,j}$ is the corresponding rating. Let $S$ be a matrix, with $S_{i,j}=g(\phi(x_i),\phi(\widebar{x}_j))$ measuring the cosine similarity between a candidate item $i$ from the victim catalogue $\mathcal{D}$ and an auxiliary item $j$ from $\widebar{\mathcal{D}}$. For each candidate user $u \in \mathcal{U}$, we compute a matching score
\begin{equation}
    \mathrm{score}(u;A_{u^\star})=\sum_{(j,y)\in A_{u^\star}}\max_{i\in\mathcal{I}: r_{u,i}=y} S_{i,j},
\end{equation}
with the corresponding term set to zero when user $u$ has no item rated as $y$. Candidate users are ranked by this score, and the true user $u^\star$ is evaluated by its rank. Similarly, we report MRR over target users, so higher values indicate that the target user is more easily re-identified.

The two right panels of Figure~\ref{fig:obfuscated_users} show user-level re-identification performance on MovieLens as the number of observed items $k$ increases (we consider the case $|\mathcal{\widebar{I}}| = 3000$). In both the closed- and open-set settings, MRR increases rapidly with $k$, indicating that a small number of rated items is often sufficient to identify the target user. Substantial reductions in user-level MRR occur only for obfuscated item representations under a title-only adversary. When the adversary uses auxiliary descriptions, however, re-identification remains effective, suggesting that semantic item similarity can link sparse user profiles even under obfuscation.

\section{Conclusion}

This work shows that natural-language item embeddings can expose substantial item-identity information in RSs settings. Even under a simple cosine-similarity attack, item identities can often be recovered with high MRR across datasets, candidate-set regimes, and embedding models. Leakage is strongest when the information available to the adversary (partially) overlaps with the victim representation, but it also persists for non-overlapping data, suggesting that identity can be encoded by semantic alignment rather than only by explicit identifiers.  We further show that this leakage can propagate to user privacy, as sparse user interactions combined with item-similarity information can support user re-identification.
Our results also highlight a leakage-utility tradeoff: obfuscating item descriptions substantially reduces leakage, but also increases rating-prediction error and reduces ranking agreement, indicating a cost to downstream recommendation quality. These findings open several directions for future work, including stronger attack models, broader domains and catalogues, and defenses that reduce identity leakage while preserving the semantic information needed for accurate recommendation.

\section*{GenAI Usage Disclosure}
Generative AI was used in the research to generate auxiliary item descriptions and obfuscated item descriptions using \texttt{gpt-oss-20b}, as described in the experimental setup. Generative AI was also used during writing only to polish grammar.

\bibliographystyle{plainnat}
\bibliography{\preprintassetpath references}

\begin{thebibliography}{41}
\providecommand{\natexlab}[1]{#1}
\providecommand{\url}[1]{\texttt{#1}}
\expandafter\ifx\csname urlstyle\endcsname\relax
  \providecommand{\doi}[1]{doi: #1}\else
  \providecommand{\doi}{doi: \begingroup \urlstyle{rm}\Url}\fi

\bibitem[Abedi~Khorasgani et~al.(2024)Abedi~Khorasgani, Mohammed, and
  Wang]{abedi2024attribute}
Hossein Abedi~Khorasgani, Noman Mohammed, and Yang Wang.
\newblock Attribute inference privacy protection for pre-trained models.
\newblock \emph{International Journal of Information Security}, 23\penalty0
  (3):\penalty0 2269--2285, 2024.

\bibitem[Alves et~al.(2021)Alves, Ledent, and Kloft]{alves2021burst}
Rodrigo Alves, Antoine Ledent, and Marius Kloft.
\newblock Burst-induced multi-armed bandit for learning recommendation.
\newblock In \emph{Proceedings of the 15th ACM Conference on Recommender
  Systems}, pages 292--301, 2021.

\bibitem[Calandrino et~al.(2011)Calandrino, Kilzer, Narayanan, Felten, and
  Shmatikov]{calandrino2011you}
Joseph~A Calandrino, Ann Kilzer, Arvind Narayanan, Edward~W Felten, and Vitaly
  Shmatikov.
\newblock " you might also like:" privacy risks of collaborative filtering.
\newblock In \emph{2011 IEEE symposium on security and privacy}, pages
  231--246. IEEE, 2011.

\bibitem[Chang et~al.(2010)Chang, Thompson, Wang, and Yao]{chang2010towards}
Chih-Cheng Chang, Brian Thompson, Hui Wang, and Danfeng Yao.
\newblock Towards publishing recommendation data with predictive anonymization.
\newblock In \emph{Proceedings of the 5th ACM Symposium on Information,
  Computer and Communications Security}, pages 24--35, 2010.

\bibitem[Chiang et~al.(2018)Chiang, Dhillon, and Hsieh]{chiang2018using}
Kai-Yang Chiang, Inderjit~S Dhillon, and Cho-Jui Hsieh.
\newblock Using side information to reliably learn low-rank matrices from
  missing and corrupted observations.
\newblock \emph{Journal of Machine Learning Research}, 19\penalty0
  (76):\penalty0 1--35, 2018.

\bibitem[Cui et~al.(2022)Cui, Ma, Zhou, Zhou, and Yang]{cui2022m6}
Zeyu Cui, Jianxin Ma, Chang Zhou, Jingren Zhou, and Hongxia Yang.
\newblock M6-rec: Generative pretrained language models are open-ended
  recommender systems.
\newblock \emph{arXiv preprint arXiv:2205.08084}, 2022.

\bibitem[Feng et~al.(2025)Feng, Li, Yu, Xiong, Fang, Zhang, Du, and
  Chen]{feng2025raid}
Xiaohua Feng, Yuyuan Li, Fengyuan Yu, Ke~Xiong, Junjie Fang, Li~Zhang, Tianyu
  Du, and Chaochao Chen.
\newblock Raid: An in-training defense against attribute inference attacks in
  recommender systems.
\newblock \emph{arXiv preprint arXiv:2504.11510}, 2025.

\bibitem[Gao et~al.(2021)Gao, Li, Feng, Chen, Zhao, He, and Jin]{gao2021cross}
Chen Gao, Yong Li, Fuli Feng, Xiangning Chen, Kai Zhao, Xiangnan He, and Depeng
  Jin.
\newblock Cross-domain recommendation with bridge-item embeddings.
\newblock \emph{ACM Transactions on Knowledge Discovery from Data (TKDD)},
  16\penalty0 (1):\penalty0 1--23, 2021.

\bibitem[Harper and Konstan(2015)]{harper2015movielens}
F~Maxwell Harper and Joseph~A Konstan.
\newblock The movielens datasets: History and context.
\newblock \emph{Acm transactions on interactive intelligent systems (tiis)},
  5\penalty0 (4):\penalty0 1--19, 2015.

\bibitem[Hou et~al.(2024)Hou, Li, He, Yan, Chen, and McAuley]{hou2024bridging}
Yupeng Hou, Jiacheng Li, Zhankui He, An~Yan, Xiusi Chen, and Julian McAuley.
\newblock Bridging language and items for retrieval and recommendation.
\newblock \emph{arXiv preprint arXiv:2403.03952}, 2024.

\bibitem[Hu and Yang(2020)]{hu2020privnet}
Guangneng Hu and Qiang Yang.
\newblock Privnet: Safeguarding private attributes in transfer learning for
  recommendation.
\newblock In \emph{Findings of the Association for Computational Linguistics:
  EMNLP 2020}, pages 4506--4516, 2020.

\bibitem[Huang et~al.(2024)Huang, Tsai, Hsiao, Lin, and
  Lin]{huang2024transferable}
Yu-Hsiang Huang, Yuche Tsai, Hsiang Hsiao, Hong-Yi Lin, and Shou-De Lin.
\newblock Transferable embedding inversion attack: Uncovering privacy risks in
  text embeddings without model queries.
\newblock In \emph{Proceedings of the 62nd Annual Meeting of the Association
  for Computational Linguistics (Volume 1: Long Papers)}, pages 4193--4205,
  2024.

\bibitem[Jeunen et~al.(2022)Jeunen, Van~Balen, and
  Goethals]{jeunen2022embarrassingly}
Olivier Jeunen, Jan Van~Balen, and Bart Goethals.
\newblock Embarrassingly shallow auto-encoders for dynamic collaborative
  filtering.
\newblock \emph{User Modeling and User-Adapted Interaction}, 32\penalty0
  (4):\penalty0 509--541, 2022.

\bibitem[Kamath(2024)]{kamath_movielens32m_sids}
Krishna Kamath.
\newblock {MovieLens-32M Movies Enriched with SIDs}.
\newblock
  \url{https://huggingface.co/datasets/krishnakamath/movielens-32m-movies-enriched-with-SIDs},
  2024.
\newblock Hugging Face dataset. Accessed: 2026-05-26.

\bibitem[Kasalick{\`y} et~al.(2025)Kasalick{\`y}, Spi{\v{s}}{\'a}k,
  Van{\v{c}}ura, Bohun{\v{e}}k, Alves, and Kord{\'\i}k]{kasalicky2025future}
Petr Kasalick{\`y}, Martin Spi{\v{s}}{\'a}k, Vojt{\v{e}}ch Van{\v{c}}ura,
  Daniel Bohun{\v{e}}k, Rodrigo Alves, and Pavel Kord{\'\i}k.
\newblock The future is sparse: Embedding compression for scalable retrieval in
  recommender systems.
\newblock In \emph{Proceedings of the Nineteenth ACM Conference on Recommender
  Systems}, pages 1099--1103, 2025.

\bibitem[Le and Abel(2025)]{le2025enhancing}
Ngoc~Luyen Le and Marie-H{\'e}l{\`e}ne Abel.
\newblock Enhancing recommender systems using textual embeddings from
  pre-trained language models.
\newblock In \emph{International Conference on Information Technology \&
  Systems}, pages 93--103. Springer, 2025.

\bibitem[Ledent et~al.(2021)Ledent, Alves, and Kloft]{ledent2021orthogonal}
Antoine Ledent, Rodrigo Alves, and Marius Kloft.
\newblock Orthogonal inductive matrix completion.
\newblock \emph{IEEE Transactions on Neural Networks and Learning Systems},
  34\penalty0 (5):\penalty0 2259--2270, 2021.

\bibitem[Li et~al.(2023)Li, Xu, and Song]{li2023sentence}
Haoran Li, Mingshi Xu, and Yangqiu Song.
\newblock Sentence embedding leaks more information than you expect: Generative
  embedding inversion attack to recover the whole sentence.
\newblock In \emph{Findings of the Association for Computational Linguistics:
  ACL 2023}, pages 14022--14040, 2023.

\bibitem[Li et~al.(2021)Li, Wang, Zhang, Ma, Cui, and Zhu]{li2021intention}
Haoyang Li, Xin Wang, Ziwei Zhang, Jianxin Ma, Peng Cui, and Wenwu Zhu.
\newblock Intention-aware sequential recommendation with structured intent
  transition.
\newblock \emph{IEEE Transactions on Knowledge and Data Engineering},
  34\penalty0 (11):\penalty0 5403--5414, 2021.

\bibitem[Li et~al.(2026)Li, Fang, Yu, Sheng, Du, Teng, Jiang, Jiang, Lin, and
  Chen]{li2026fedau2}
Yuyuan Li, Junjie Fang, Fengyuan Yu, Xichun Sheng, Tianyu Du, Xuyang Teng,
  Shaowei Jiang, Linbo Jiang, Jianan Lin, and Chaochao Chen.
\newblock Fedau2: Attribute unlearning for user-level federated recommender
  systems with adaptive and robust adversarial training.
\newblock In \emph{Proceedings of the AAAI Conference on Artificial
  Intelligence}, volume~40, pages 23310--23318, 2026.

\bibitem[Mahloujifar et~al.(2021)Mahloujifar, Inan, Chase, Ghosh, and
  Hasegawa]{mahloujifar2021membership}
Saeed Mahloujifar, Huseyin~A Inan, Melissa Chase, Esha Ghosh, and Marcello
  Hasegawa.
\newblock Membership inference on word embedding and beyond (2021).
\newblock \emph{arXiv preprint arXiv:2106.11384}, 2021.

\bibitem[Mateos and Bellog{\'\i}n(2024)]{mateos2024systematic}
Pablo Mateos and Alejandro Bellog{\'\i}n.
\newblock A systematic literature review of recent advances on context-aware
  recommender systems.
\newblock \emph{Artificial Intelligence Review}, 58\penalty0 (1):\penalty0 20,
  2024.

\bibitem[Mattern et~al.(2023)Mattern, Mireshghallah, Jin, Sch{\"o}lkopf,
  Sachan, and Berg-Kirkpatrick]{mattern2023membership}
Justus Mattern, Fatemehsadat Mireshghallah, Zhijing Jin, Bernhard
  Sch{\"o}lkopf, Mrinmaya Sachan, and Taylor Berg-Kirkpatrick.
\newblock Membership inference attacks against language models via
  neighbourhood comparison.
\newblock In \emph{Findings of the Association for Computational Linguistics:
  ACL 2023}, pages 11330--11343, 2023.

\bibitem[Morris et~al.(2023)Morris, Kuleshov, Shmatikov, and
  Rush]{morris2023text}
John Morris, Volodymyr Kuleshov, Vitaly Shmatikov, and Alexander~M Rush.
\newblock Text embeddings reveal (almost) as much as text.
\newblock In \emph{Proceedings of the 2023 Conference on Empirical Methods in
  Natural Language Processing}, pages 12448--12460, 2023.

\bibitem[Narayanan and Shmatikov(2008)]{narayanan2008robust}
Arvind Narayanan and Vitaly Shmatikov.
\newblock Robust de-anonymization of large sparse datasets.
\newblock In \emph{2008 IEEE Symposium on Security and Privacy (sp 2008)},
  pages 111--125. IEEE, 2008.

\bibitem[Ni et~al.(2026)Ni, Yuanpeng, Zhou, Hong, Zhang, Cai, Wu, Zeng, Yu, and
  Li]{ni2026treebridge}
Yabo Ni, Cao Yuanpeng, Wenhang Zhou, Bangyang Hong, Zhongyi Zhang, Enlei Cai,
  Kangle Wu, Anxiang Zeng, Han Yu, and Xiaoxiao Li.
\newblock Treebridge: Aligning llm embeddings in industrial recommender
  systems.
\newblock In \emph{Proceedings of the AAAI Conference on Artificial
  Intelligence}, volume~40, pages 40370--40377, 2026.

\bibitem[Oghina et~al.(2012)Oghina, Breuss, Tsagkias, and
  De~Rijke]{oghina2012predicting}
Andrei Oghina, Mathias Breuss, Manos Tsagkias, and Maarten De~Rijke.
\newblock Predicting imdb movie ratings using social media.
\newblock In \emph{European conference on information retrieval}, pages
  503--507. Springer, 2012.

\bibitem[Sayifullah et~al.(2026)Sayifullah, Nadifa, Musa, Dongka, irawaty
  Tolango, and Bonok]{sayifullah2026exploratory}
M~Fauzan Sayifullah, Ulfatun Nadifa, Wahab Musa, Rahmat~Hidayat Dongka, Ade
  irawaty Tolango, and Zainudin Bonok.
\newblock Exploratory data analysis on tmdb top rated movies dataset using
  winsorization approach.
\newblock \emph{International Journal of Embedded Computer Engineering},
  1\penalty0 (1):\penalty0 18--25, 2026.

\bibitem[Song and Raghunathan(2020)]{song2020information}
Congzheng Song and Ananth Raghunathan.
\newblock Information leakage in embedding models.
\newblock In \emph{Proceedings of the 2020 ACM SIGSAC conference on computer
  and communications security}, pages 377--390, 2020.

\bibitem[Song and Shmatikov(2019)]{song2019overlearning}
Congzheng Song and Vitaly Shmatikov.
\newblock Overlearning reveals sensitive attributes.
\newblock \emph{arXiv preprint arXiv:1905.11742}, 2019.

\bibitem[Torres and Olivares(2024)]{torres2024anonymizing}
Nicol{\'a}s Torres and Patricio Olivares.
\newblock De-anonymizing users across rating datasets via record linkage and
  quasi-identifier attacks.
\newblock \emph{Data}, 9\penalty0 (6):\penalty0 75, 2024.

\bibitem[Vats et~al.(2025)Vats, Raja, Jain, and Chadha]{vats2025exploring}
Arpita Vats, Rahul Raja, Vinija Jain, and Aman Chadha.
\newblock Exploring the impact of large language models on recommender systems:
  An extensive review.
\newblock In \emph{2025 IEEE International Conference on Big Data (BigData)},
  pages 3081--3090. IEEE, 2025.

\bibitem[Wan and McAuley(2018)]{wan2018item}
Mengting Wan and Julian McAuley.
\newblock Item recommendation on monotonic behavior chains.
\newblock In \emph{Proceedings of the 12th ACM conference on recommender
  systems}, pages 86--94, 2018.

\bibitem[Weinsberg et~al.(2012)Weinsberg, Bhagat, Ioannidis, and
  Taft]{weinsberg2012blurme}
Udi Weinsberg, Smriti Bhagat, Stratis Ioannidis, and Nina Taft.
\newblock Blurme: Inferring and obfuscating user gender based on ratings.
\newblock In \emph{Proceedings of the sixth ACM conference on Recommender
  systems}, pages 195--202, 2012.

\bibitem[Wu et~al.(2020)Wu, Qiao, Chen, Wu, Qi, Lian, Liu, Xie, Gao, Wu, and
  Zhou]{wu2020mind}
Fangzhao Wu, Ying Qiao, Jiun-Hung Chen, Chuhan Wu, Tao Qi, Jianxun Lian,
  Danyang Liu, Xing Xie, Jianfeng Gao, Winnie Wu, and Ming Zhou.
\newblock Mind: A large-scale dataset for news recommendation.
\newblock In \emph{Proceedings of the 58th Annual Meeting of the Association
  for Computational Linguistics}, pages 3597--3606, 2020.

\bibitem[Wu et~al.(2025)Wu, Jiang, and Hu]{wu2025aegis}
Wenhan Wu, Jiawei Jiang, and Chuang Hu.
\newblock Aegis: Post-training attribute unlearning in federated recommender
  systems against attribute inference attacks.
\newblock In \emph{Proceedings of the ACM on Web Conference 2025}, pages
  3783--3793, 2025.

\bibitem[Zhang et~al.(2021{\natexlab{a}})Zhang, Ren, Wang, Ren, Chen, Hu, and
  Zhang]{zhang2021membership}
Minxing Zhang, Zhaochun Ren, Zihan Wang, Pengjie Ren, Zhunmin Chen, Pengfei Hu,
  and Yang Zhang.
\newblock Membership inference attacks against recommender systems.
\newblock In \emph{Proceedings of the 2021 ACM SIGSAC Conference on Computer
  and Communications Security}, pages 864--879, 2021{\natexlab{a}}.

\bibitem[Zhang et~al.(2021{\natexlab{b}})Zhang, Yin, Chen, Huang, Cui, and
  Zhang]{zhang2021graph}
Shijie Zhang, Hongzhi Yin, Tong Chen, Zi~Huang, Lizhen Cui, and Xiangliang
  Zhang.
\newblock Graph embedding for recommendation against attribute inference
  attacks.
\newblock In \emph{Proceedings of the Web Conference 2021}, pages 3002--3014,
  2021{\natexlab{b}}.

\bibitem[Zhao et~al.(2024)Zhao, Fan, Li, Liu, Mei, Wang, Wen, Wang, Zhao, Tang,
  et~al.]{zhao2024recommender}
Zihuai Zhao, Wenqi Fan, Jiatong Li, Yunqing Liu, Xiaowei Mei, Yiqi Wang, Zhen
  Wen, Fei Wang, Xiangyu Zhao, Jiliang Tang, et~al.
\newblock Recommender systems in the era of large language models (llms).
\newblock \emph{IEEE Transactions on Knowledge and Data Engineering},
  36\penalty0 (11):\penalty0 6889--6907, 2024.

\bibitem[Zhong et~al.(2024)Zhong, Wang, Xu, Xu, and Wang]{zhong2024interaction}
Da~Zhong, Xiuling Wang, Zhichao Xu, Jun Xu, and Wendy~Hui Wang.
\newblock Interaction-level membership inference attack against recommender
  systems with long-tailed distribution.
\newblock In \emph{Proceedings of the 33rd ACM International Conference on
  Information and Knowledge Management}, pages 3433--3442, 2024.

\bibitem[Zhu et~al.(2024)Zhu, Shao, Lian, Wu, Liu, Yang, and
  Chen]{zhu2024understanding}
Zhihao Zhu, Ninglu Shao, Defu Lian, Chenwang Wu, Zheng Liu, Yi~Yang, and Enhong
  Chen.
\newblock Understanding privacy risks of embeddings induced by large language
  models.
\newblock \emph{arXiv preprint arXiv:2404.16587}, 2024.

\end{thebibliography}

\end{document}